\documentclass[%
superscriptaddress,
reprint,
 amsmath,amssymb,
 aps,
 pra,
]{revtex4-1}

\usepackage{graphicx}% Include figure files
\usepackage{dcolumn}% Align table columns on decimal point
\usepackage{bm}% bold math
\usepackage{mathtools}
\usepackage{amsmath}
\usepackage{appendix}

\newcommand{\vv}[1]{\mathbf{#1}}
\usepackage{hyperref}% add hypertext capabilities
\begin{document}

\preprint{APS/123-QED}

\title{Using Machine Learning to Augment Coarse-Grid Computational Fluid Dynamics Simulations\\}% Force line breaks with \\

\author{Jaideep Pathak}
\email{Corresponding Author: jpathak@lbl.gov}
\affiliation{NERSC, Lawrence Berkeley National Laboratory}
\author{Mustafa Mustafa}
\altaffiliation{Jaideep Pathak and Mustafa Mustafa contributed equally to this paper}
\affiliation{NERSC, Lawrence Berkeley National Laboratory}
\altaffiliation{Jaideep Pathak and Mustafa Mustafa contributed equally to this paper}
\author{Karthik Kashinath}
\affiliation{NERSC, Lawrence Berkeley National Laboratory}
\author{Emmanuel Motheau}
\affiliation{CRD, Lawrence Berkeley National Laboratory}
\author{Thorsten Kurth}
\affiliation{Nvidia Corporation}
\author{Marcus Day}
\affiliation{CRD, Lawrence Berkeley National Laboratory}

\date{\today}

\begin{abstract}
Simulation of turbulent flows at high Reynolds number is a computationally challenging task relevant to a large number of engineering and scientific applications in diverse fields such as climate science, aerodynamics, and combustion. Turbulent flows are typically modeled by the Navier-Stokes equations. Direct Numerical Simulation (DNS) of the Navier-Stokes equations with sufficient numerical resolution to capture all the relevant scales of the turbulent motions can be prohibitively expensive. Simulation at lower-resolution on a coarse-grid introduces significant errors. We introduce a machine learning (ML) technique based on a deep neural network architecture that corrects  the numerical errors induced by a coarse-grid simulation of turbulent flows at high-Reynolds numbers, while simultaneously recovering an estimate of the high-resolution fields. Our proposed simulation strategy is a hybrid ML-PDE solver that is capable of obtaining a meaningful high-resolution solution trajectory while solving the system PDE at a lower resolution.  The approach has the potential to dramatically reduce the expense of turbulent flow simulations. As a proof-of-concept, we demonstrate our ML-PDE strategy on a two-dimensional turbulent (Rayleigh Number  $Ra=10^9$) Rayleigh-B{\' e}nard Convection (RBC) problem.
\end{abstract}

\maketitle

Most practical flows of interest are by nature turbulent and present a wide range of temporal and spatial scales.  The modeling of such applications is challenging and a Direct Numerical Simulation (DNS) approach, in which the full range of spatial and temporal scales exhibited by the governing equations are resolved numerically, imposes a severe computational burden. The Reynolds number is a non-dimensional measure of the range of temporal and spatial scales present in a system~\cite{grossmann2002prandtl}~\footnote{Note that in thermal convection of the form considered in this paper, the Reynolds number has been shown to depend on the Rayleigh number as an approximate power law of the form $Re \sim Ra^{\beta}$ for a fixed Prandtl number. c.f. Ref [1]}, and thus plays a key role in determining the computational resources required to accurately simulate a flow system. One of the first DNS studies of turbulence was limited to a Reynolds number $Re \approx 500$ \cite{Moser:1999}, whereas about $15$ years later, the same group reported a similar study \cite{Lee:2015} with $Re \approx 5000$ -- just one order of magnitude higher. Consequently, DNS of much higher Reynolds number flows present in practical engineering applications will likely remain out of reach for some time.

Several recent papers have demonstrated the effective incorporation of machine learning (ML) for applications in fluid modeling, from augmenting Reynolds Averaged Navier-Stokes (RANS) solvers and ML-based subgrid-scale closure models to fully data-driven ML fluid solvers~\cite{brunton2020machine, duraisamy2019turbulence, maulik2019subgrid, pathak2018model, rasp2018deep}. However, the computational cost of training pure ML CFD solvers remains prohibitively high for all but the simplest of problems.
While there has been a number of impressive advances in the development of data-driven models of the Earth's atmosphere~\cite{scher2018toward, scher2019weather, dueben2018challenges, weyn2019can, arcomano2020machine}, for example, these models are considerably more expensive to run than state-of-the-art physics based General Circulation Models with comparable accuracy. Data availability and computational resource constraints thus warrant the development of a hybrid approach where a physics-based numerical solver works in tandem with a coupled ML architecture, thus leveraging the strength of each approach while maintaining a reasonable computational cost. Such hybrid approaches for model error correction have been previously demonstrated on low-dimensional chaotic systems~\cite{pathak2018hybrid, wan2018data}. Data-driven PDE discretizations for fluid flow were discussed in Refs.~\cite{bar2019learning, zhuang2020learned}. Recent advances in Single Image Super-Resolution have demonstrated impressive results in the field of computer vision and image processing (c.f. Ref.~\cite{yang2019deep} for a review), where deep neural networks are used to increase the resolution of coarse-grained images. Super-Resolution techniques have been recently demonstrated for fluid dynamics and climate applications~\cite{jiang2020meshfreeflownet, stengel2020adversarial, liu2020deep, xie2018tempogan}.

The main idea of this paper is to develop a strategy based on Deep Learning (DL) methods that has the potential to extend the Reynolds numbers accessible to detailed simulations. The present work demonstrates an example flow simulation computed on a coarse mesh that is enhanced using a DL model to populate the finer scales that are normally available only by increasing resolution, and expense, of the simulation. This technique also introduces a correction to the model errors resulting from the simulating on a coarse mesh. In contrast to a \textit{post-facto} Super-Resolution applications that works with artificially coarsened simulation data, we present a general technique to enhance PDE simulation data that is generated by the flow solver at low resolution.  Our approach results in a high-resolution estimate of system variables while simultaneously correcting model error introduced during the coarse-grid PDE simulation.

The remainder of this paper is structured as follows. In section~\ref{sec:methodology}, the basic computational methodology is presented. Next, in section~\ref{sec:rbc}, the canonical problem of Rayleigh-B{\' e}nard Convection (RBC) in two dimensions is reviewed. In section~\ref{sec:mlpde}, our novel Deep Learning correction algorithm method is presented. Finally, in section~\ref{sec:results}  the results show that it is possible to capture small physical details with a coarse simulation that is coupled with a Deep Learning algorithm, and that important temporal and spectral properties of the flow can be recovered that compare well to a \emph{ground truth} simulation at higher resolution.

\section{\label{sec:methodology}Methodology}

In this section we formalize the problem definition and proposed solution. We also state the goal of this paper in plain language. The main challenge is that the trajectory of a coarse-grid simulation can be very different from that of a well-resolved solution. We investigate whether it is possible to compute on-the-fly corrections to the coarse-grid trajectory so that they then follow that of the fine grid. We do this by constructing an ML model that: 1) learns to model the error on the large scales due to the small scales that are missing from the coarse simulation, and 2) populates the missing small scales at each step of the low resolution solver.

\subsection{\label{sec:problem} Problem Definition}

Consider a physical system whose evolution is described by a set of PDEs (such as the Navier-Stokes equations). Let the state of the system at time $t$ be denoted by $\vv{\Psi}(t)$. Typically, $\vv{\Psi}(t)$ will be a multi-channel tensor representing the field of a physical variable such as the temperature, pressure and velocity components. The evolution of $\vv{\Psi}(t)$ under the dynamics of the PDE can be represented by the initial value problem:
\begin{align}\label{eq:abstracteqn}
\partial_t{\vv{\Psi}}(t) &= \vv{ \mathcal{F}}[\vv{\Psi}(t), \partial_x \vv{\Psi}(t)].
\end{align}
with initial conditions, $\vv{\Psi}(t=0)$. In Eq.~(\ref{eq:abstracteqn}), $\mathcal{F}$ denotes a set of operators that act on $\vv{\Psi}$ and its set of (first or higher order) spatial derivatives, denoted by $\partial_x \vv{\Psi}(t)$. 
\\

A variety of techniques may be employed to numerically evolve Eq.~(\ref{eq:abstracteqn}) in time, based on numerical resolution parameterized by the integer tuple, $N$. For finite-difference or finite-element approaches, $N$=$(N_1,\ldots,N_D)$, would represent the number of mesh points across the domain in each direction, $D$; for spectral solvers, $N$ might specify the number of corresponding Fourier modes. Assume we have such a solver, represented by the nonlinear operator $\vv{F}_{N}$. The operator $\vv{F}_{N}$ acts on the fields $\vv{X}_{N}(t)$ to evolve them in time by a time interval $\delta t_{N}$, as an approximation solution to Eq.~(\ref{eq:abstracteqn}), subject to initial conditions, $\vv{X}_{N}(t_0) = \vv{X}_{N}^0$. Note that $\delta t_{N}$ typically is smaller for increasing $N$.

We are interested in the value of the field $\vv{X}_{N} (t_0 + T)$ at some later time, $t_0 + T$. We apply the operator $\vv{F}_{N}$ on the fields $\vv{X}(t)$ so that,
\begin{align}\label{eq:solver_hr}
\vv{X}_{N}(t_0 + T) &=  \vv{F}_{N}^{(T)}[\vv{X}_{N}(t_0)].
\end{align}
Here, $\vv{F}^{(T)}_{N}$ simply denotes the composite operator that evolves the fields over an interval $T$ via a sequence of multiple (perhaps variable sized) time steps. To save computational cost, we could also choose to evolve the appropriately coarsened initial condition at a lower resolution  $N^\prime$=$(N_1/m, \ldots, N_D/m)$ using the operator denoted $\vv{F}^{\prime (T)}_{N^{\prime}}$. To coarsen the initial condition, we interpolate the fields $\vv{X}_{N}(t_0)$ onto the coarser grid with resolution, $N^\prime$. We call this operation `down-scaling' and denote the down-scaling interpolation operator by $\mathcal{D}_m$. The appropriate form of the operator $\mathcal{D}_m$ may depend on the type of CFD solver and the nature of the fields being interpolated, among other factors. We also construct a corresponding `up-scaling' operator $\mathcal{U}_m$, that transforms low-resolution fields into high-resolution. Note that we do not assume $\mathcal{U}_m$ to be particularly sophisticated. A simple example of an up-scaling operator could be one that pads pixels with the value of nearest neighbors. 
In general, due to the highly nonlinear interactions across length and time scales, 
\begin{align}
 \vv{F}^{(T)} \vv{X}(t_0)   \neq \mathcal{U}_m \vv{F}^{\prime (T)} \left[  \mathcal{D}_m \vv{X}(t_0)  \right].
\end{align}
Because information is lost in the down-scaling operation, it is not generally possible to recover the output of a high-resolution PDE solver from a low-resolution simulation over a finite time interval. However, we posit that for an interval of time $\tau$ that is very small compared to characteristic macroscopic time-scales of the system (such as the largest eddy turnover time), 
\begin{align}
\vv{F}^{(\tau)} \left[ \vv{X}(t_0) \right]  = \mathcal{U}_m \vv{F}^{\prime (\tau)} \left[  \mathcal{D}_m \vv{X}(t_0)  \right] + \epsilon.
\end{align}
If we are able to model the error $\vv{\epsilon}$ then we can estimate and correct for the model error at regular intervals, $\tau$, and estimate a corrected trajectory $\vv{X}^{ml}(t)$ that is close to the true high-resolution trajectory $\vv{X}(t)$.

\subsection{\label{sec:ml} Model Error correction with Machine Learning}

We consider a Machine Learning technique to correct the model error in a low-resolution PDE simulation and simultaneously recover the high-resolution fields.
\begin{figure*}
\centering
\includegraphics[width =0.7 \textwidth]{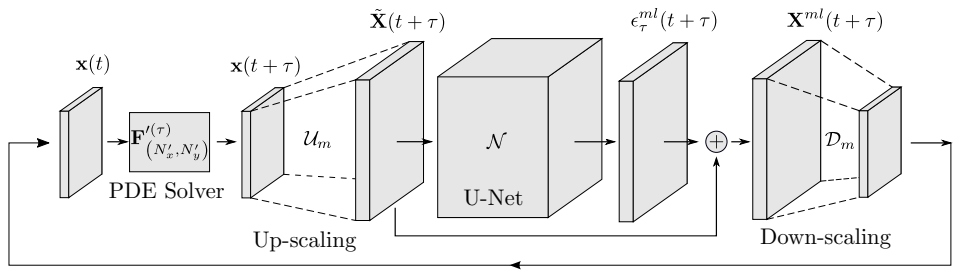}
\caption{MLPDE hybrid architecture illustrating the algorithm given by Eq.~(\ref{eq:inference}). The low resolution PDE time-stepper is followed by naive up-scaling, correction using a deep neural network and down-scaling with the process repeating in a closed feedback loop.  }
\label{fig:sr_arch}
\end{figure*}
We choose a small trajectory correction time interval, $\tau$, and emphasize that over this interval the CFD solver may take multiple (possibly adaptively chosen) time steps (typically chosen for stability/accuracy of the numerical integration scheme). We seek to compute model errors at regular intervals of $\tau$ in order to correct the PDE trajectory computed at low resolution using ML, and also to estimate the missing high-resolution fields. We train a supervised neural network to model the $\tau$-step error field $\vv{\epsilon}_{\tau}(t)$ which is defined  by the following equations.

\begin{align}
\vv{X}(t +  \tau) &=  \vv{F}_{N}^{(\tau)} \left[ \vv{X}(t) \right], \label{eq:sr1}\\
\vv{x}(t + \tau) &=  \vv{F}_{N^\prime}^{\prime (\tau)} \left[ \mathcal{D}_m \left[ \vv{X}(t)\right] \right], \label{eq:sr2}\\
\tilde{\vv{X} }(t + \tau) &= \mathcal{U}_m\left[ \vv{x}(t + \tau) \right], \label{eq:sr3} \\
\vv{\epsilon}_{\tau }(t + \tau) &= \vv{X}(t + \tau) - \tilde{\vv{X} }(t + \tau). \label{eq:sr4}
\end{align}
The supervised neural network, denoted by $\mathcal{N}$ is trained to obtain an estimate $\epsilon^{ml}_{\tau}(t)$ of the model error $\epsilon_{\tau}(t)$
\begin{align}
\vv{\epsilon}^{ml}_{\tau}(t) &= \mathcal{N}\left[ \tilde{\vv{X}}(t) \right].
\end{align}
We operate the hybrid ML-PDE solver in inference mode as follows
\begin{itemize}
\item \textit{Initialize}: Start from the initial condition $\vv{X}_{N}(t_0)$ and initialize the ML-estimated trajectory $\vv{X}^{ml}(t_0)$ so that
\begin{align}
\vv{X}^{ml}(t_0) &= \vv{X}_{N}(t_0).
\end{align}
\item \textit{Timestep with corrections}: The following equations are computed in a loop:
\begin{align}
\vv{x}(t +  \tau) &=  \vv{F}_{N^\prime}^{\prime (\tau)}  \left[ \mathcal{D}_m \left [ \vv{X}^{ml}(t) \right]\right], \label{eq:timestep_lr} \\ 
\tilde{\vv{X}}(t + \tau) &= \mathcal{U}_m\left[ \vv{x}(t + \tau \right)], \label{eq:bicubic} \\
\epsilon^{ml}_\tau(t+\tau) &= \mathcal{N}\left[ \tilde{\vv{X}}(t + \tau) \right], \label{eq:nn_correction}\\
\vv{X}^{ml}(t+\tau) &= \tilde{\vv{X}}(t + \tau) + \epsilon^{ml}_\tau(t+\tau). \label{eq:apply_correction}
\end{align}
Eqs.~(\ref{eq:timestep_lr} - \ref{eq:apply_correction}) above can be combined to give us a single inference equation as follows:
\begin{multline}\label{eq:inference}
\vv{X}^{ml}(t+\tau) = \mathcal{U}_m \left[  \vv{F}_{N^\prime}^{\prime (\tau)}  \left[ \mathcal{D}_m \left [ \vv{X}^{ml}(t) \right]\right]  \right] + \\  \mathcal{N}\left[  \mathcal{U}_m \left[  \vv{F}_{N^\prime}^{\prime (\tau)}  \left[ \mathcal{D}_m \left [ \vv{X}^{ml}(t) \right]\right]  \right]  \right].
\end{multline}
\end{itemize}

This completes our problem definition and proposed framework for coupling Machine Learning with a PDE solver. In the next two sections we turn our focus to a concrete implementation of an ML architecture which we couple to a solver to demonstrate this framework for solving a canonical 2-dimensional fluid convection problem, namely the Rayleigh-B{\'e}nard system of equations.

\section{\label{sec:rbc} Rayleigh-B{\' e}nard Convection (RBC) }

In order to demonstrate the effectiveness of our hybrid ML-PDE architecture, we consider a two-dimensional Rayleigh-B{\' e}nard Convection (RBC) problem operating in a regime that exhibits moderate levels of fine-scale turbulent fluctuations.  The RBC problem is modeled with the incompressible Navier-Stokes equations formulated under the Boussinesq approximation. Nondimensionalization by the Rayleigh and Prandtl numbers and subtracting the steady conduction-only solution, gives the following formulation:
\begin{align}
\quad \nabla \cdot \mathbf{u} &=0 , \label{eq:rbc1} \\
\partial_{t} \mathbf{u} &= - \left(\mathbf{u} \cdot \nabla\right) \mathbf{u}  -\nabla p+\sqrt{\frac{P r}{R a }} \nabla^{2} \mathbf{u}+ \theta \mathbf{e}_{z} , \label{eq:rbc2}\\
\partial_{t} \theta  &=\sqrt{\frac{1}{\operatorname{Pr} R a}} \nabla^{2} \theta  - \left(\mathbf{u} \cdot \nabla\right) \theta + \mathbf{u} \cdot \mathbf{e}_{z}. \label{eq:rbc3}
\end{align}
where $\theta$ and $p$ are the (nondimensional) deviations of temperature and pressure from the steady solution, and $\mathbf{u}$ is the nondimensionalized fluid velocity. The detailed derivation of this set of equations and the nondimensionalization parameters can be found in \cite{Pandey:2016}. In the present study, the Prandtl and Rayleigh numbers are set to $Pr=0.7$ and  $Ra=10^9$, respectively.

The equations~(\ref{eq:rbc1})-(\ref{eq:rbc3}) are solved in a 2D computational domain with unit aspect ratio, $\Gamma$=$1$. No-slip ($\mathbf{u}$=$0$), isothermal ($\theta$=$0$) boundary conditions are imposed on the upper and lower walls, while periodicity is applied to the lateral boundaries. The initial velocity and pressure fluctuations are set to zero and an initial profile on the fluctuating temperature is created with a random seed.

To evolve the system numerically, we use Dedalus~\cite{2020PhRvR...2b3068B}, an open-source spectral framework for solving a general set of partial differential equations. We note that due to the non-dissipative nature of the spectral discretization used in Dedalus, stability considerations limit the maximum numerical time step as a function of $Ra$. In the cases considered below, the solutions were computed on the coarsest meshes using 1/10 of the computed stable step size. Larger values of $Ra$ would increase the energy contained within the under-resolved fluctuations at high wave number but would require an even further reduced time step for stable temporal evolution.

\section{\label{sec:mlpde} ML-PDE architecture}

In this section we describe the hybrid ML-PDE architecture that combines a PDE time stepper operating at a coarse-resolution with a convolutional ML architecture operating as a model error corrector. The hybrid architecture has a training phase and an inference phase as we outline below. We also outline the schemes used for up- and down-scaling.

\subsection{\label{sec:interpolation} Up-scaling and down-scaling Operators}
Throughout the rest of this paper, we will often need to transform fields from high-resolution to low resolution and vice-versa.
%as described in Eqs.~(\ref{eq:downsample}, \ref{eq:upsample}).
Since we are using a spectral CFD solver a natural choice for the up-scaling and down-scaling operators is one that pads or truncates modes in spectral space. Dedalus provides the functionality to implement such transforms natively. When downsampling a field with $N_s$ spectral coefficients by a factor of $m$, the $N_s$ spectral coefficients are truncated after the first $m$ modes and are then transformed to a $1/m$ times scaled grid in real space. When upsampling, the spectral coefficients are padded with the appropriate number of zeros above the highest modes before transforming to an $m$ times scaled grid in real space. We denote these spectral down-scaling and up-scaling operators by $\vv{D}_m$ and $\vv{U}_m$ respectively.

\subsection{\label{sec:training} Training}

\subsubsection{\label{sec:data} High-Resolution Data Generation}
Training data was generated using the Dedalus PDE solver to simulate Eqs.~(\ref{eq:rbc1}-\ref{eq:rbc3}) on a Cartesian, evenly spaced grid with resolution of $(N_{x}, N_{z})$=$(512, 512)$. After the system reached quasi-steady conditions, we saved snapshots of the state at intervals of $\tau = 0.05$ Model Time Units (MTU). For comparison, the largest eddy turnover time was estimated to be approximately $2.5$ MTU. We create a 4-channel tensor $\vv{X}_{512}(t_k)$ by stacking the instantaneous velocity, temperature and pressure fields at each spatial index. These high-resolution fields $\vv{X}_{512}(t_k)$ will be referred to as the \textit{ground truth} fields.

\subsubsection{\label{sec:regrid} Regridding}
For every saved snapshot $\vv{X}_{512}(t_k)$, we generate a low-resolution down-scaled snapshot  $\vv{x}_{128}(t_k) = \vv{D}_4 \vv{X}_{512}(t_k)$ on a Cartesian, evenly spaced grid with resolution $(N_x, N_z) = (128, 128) $, using the down-scaling operator, $\vv{D}_m$, described in section~\ref{sec:interpolation} with $m = 4$.

\subsubsection{\label{sec:pairs} Pair Creation}
We solve Eqs.~(\ref{eq:rbc1} - \ref{eq:rbc3}) using Dedalus on a $128 \times 128$ grid to evolve each of the regridded snapshots $\vv{x}_{128}(t_k)$ by a time interval $\tau$. We denote the time-stepped snapshot at time $t = t_k + \tau$ by $\tilde{\mathbf{x }}_{128}(t_{k + 1})$. We then up-scale $\tilde{\vv{x}}_{128}(t_{k+1})$ using $\vv{U}_4$ to obtain the interpolated tensor $\tilde{\vv{X}}_{512}(t_{k+1})$. The high-resolution \textit{ground truth} field at time $t = t_k + \tau$, $\vv{X}_{512}(t_{k+1})$, is then used to compute the error tensor, $\vv{E}(t_{k+1}) \coloneqq (\vv{X}_{512}(t_{k+1}) - \tilde{\vv{X}}_{512}(t_{k+1}) $. An ordered pair $\vv{P}_{k+1}$ is then defined according to:
\begin{align}
\vv{P}_{k + 1}  \coloneqq \left( \tilde{\vv{X}}_{512}(t_{k+1}) , \vv{E}(t_{k+1} )  \right).
\end{align}

\subsubsection{\label{sec:unet} ML model training}
We use $N_{train} = 10^5$ pairs $\lbrace \vv{P}_k \rbrace_{k=1}^{10^5}$ as input output pairs to train a deep Convolutional Neural Network (CNN) with a UNet~\cite{ronneberger2015u} style architecture. We further use $N_{validate} = 3 \times 10^4$ pairs to tune the network architecture and optimize hyper-parameters. The validation dataset was also used to monitor the validation loss during training the network, however, a separate test dataset was used for the final performance evaluation of the ML-PDE framework. The details of the architecture and tuned hyper-parameters are detailed in Appendix \ref{appendix:NN}. The network is trained with an L1 Norm loss between the network prediction and the ground truth fields.

The all-convolutional network architecture is independent of the input spatial grid size, this allows us to train the network on smaller grid inputs and then apply it on the full grid during inference. We leverage this feature to train the network on random $(256 \times 256)$ on-the-fly crops of the full $(512 \times 512)$ input fields. In addition to allowing us to train the network faster, this also assures us that the network uses relatively local information to estimate the missing finer scales. The success of this strategy implies that the learned error model dynamically adapts to the scales and structures in the input field, for example, near the center of the turbulence versus near the walls. This strategy also opens the door for building constant network-size error models that generalize well to larger grids, a promising strategy for applications to 3D fields.

For implementation, we use PyTorch~\cite{NEURIPS2019_9015} to build the neural network and RayTune~\cite{moritz2018ray, liaw2018tune} for distributed hyper-parameter optimization. The final network was trained with a distributed data-parallel strategy on 6 NVIDIA V100 GPUs on the GPU partition of NERSC Cori HPC system. We use PyTorch's Distributed Data Parallel package~\cite{torch_ddp} for the distributed training while leveraging NVIDIA's Collective Communications Library (NCCL)~\cite{nccl} backend for best scaling performance on multi-GPUs. We also use PyTorch Automatic Mixed Precision (AMP) API~\cite{micikevicius2018mixed, torch_amp} during network training, this reduces GPU memory utilization which allowed us to train a bigger model. AMP leverages the Tensor Cores on the V100 GPUs which resulted in 2.5x speed-up during network training. Finally, we made extensive use of Weights and Biases~\cite{wandb} for tracking experiments.

\subsection{Inference}
We generate $10^4$ snapshots $\vv{X}_{512}(t_k)$ by initializing the solver with a new initial condition (separate from all initial conditions used to create the training and validation data sets) using the Dedalus PDE solver at resolution $512 \times 512$. From this test set, we pick 20 random snapshots denoted $\lbrace \vv{X}_{512}^{i} \rbrace_{i=1}^{20}$. From each of these snapshots, we generate a down-scaled snapshot $\lbrace \vv{x}_{128}^i \rbrace_{i=1}^{20}$ using the procedure described in Sec.~\ref{sec:regrid}.

\paragraph*{Ground Truth:}
The trajectories resulting from evolving each of the the initial conditions in $\lbrace \vv{X}_{512}^{i}\rbrace_{i=1}^{20}$ using Eqs.~(\ref{eq:rbc1} - \ref{eq:rbc3}) with Dedalus at resolution $512 \times 512$ will be considered the `ground truth' and denoted by $\lbrace \vv{X}^{i}_{truth}(t) \rbrace$

\paragraph*{Baseline:}
As a baseline comparison, we evolve each of the $\lbrace \vv{x}_{128}^i \rbrace_{i=1}^{20}$ using Eqs.~(\ref{eq:rbc1} - \ref{eq:rbc3}) at $128 \times 128$ resolution using Dedalus and denote the resulting trajectories by $\vv{x}^{i}_{base}(t)$

\paragraph*{ML:}
Starting with each of the initial conditions in $\lbrace \vv{X}_{512}^{i}\rbrace_{i=1}^{20}$ we use the trained UNet in conjunction with the Dedalus PDE solver at resolution $128 \times 128$ to generate a trajectory according to Eq.~(\ref{eq:inference}). The resulting trajectory is denoted by $\lbrace \vv{X}^{i}_{ml}(t) \rbrace_{i=1}^{20}$.

\section{\label{sec:results} Results}

\begin{figure}
\includegraphics[width = 0.4\textwidth]{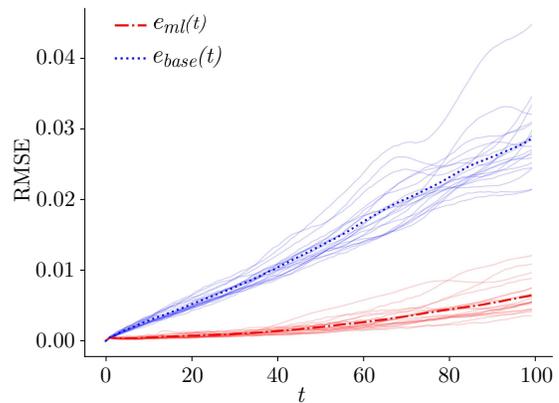}
\caption{The RMS error in the ML augmented simulation $e^{i}_{ml}(t)$ (red) and the Baseline low resolution PDE simulation $e^i_{base}(t)$ (blue) for 20 different trajectories starting from differing initial conditions. The dark red (blue) line indicates the average RMS error over the 20 trajectories for the ML augmented (baseline) simulation}
\label{fig:rms}
\end{figure}

We evaluate the fidelity of the hybrid ML-PDE trajectories to the ground truth trajectories by evaluating the Root Mean Square (RMS) error $e^{i}_{ml}(t) = \langle \lVert \vv{D}_4\vv{X}^{i}_{ml}(t)- \vv{D}_4 \vv{X}_{truth}^{i}(t) \rVert^2\rangle^{1/2} $. Additionally we also evaluate the baseline RMS error $e^i_{base}(t) = \langle \lVert  \vv{x}^{i}_{base}(t)- \vv{D}_4 \left[\vv{X}_{truth}^{i}(t)  \right]  \rVert^2 \rangle^{1/2}$. Fig.~\ref{fig:rms} shows the RMS error curves $e^i_{ml}$, $e^i_{base}$ for the 20 trials along with the mean RMS error over the 20 trials.

\begin{figure}
\includegraphics[width = 0.45\textwidth]{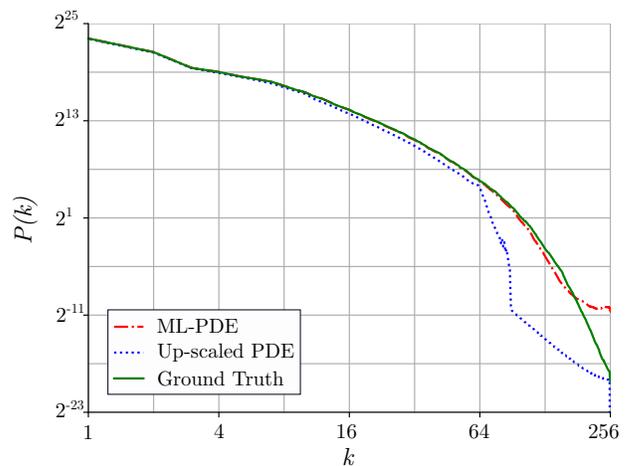}
\caption{The Power Spectral Density (PSD) of the $\theta$ variable in the ML-PDE trajectory $\vv{X}^{i}_{ml}(t)$ (red), the ground truth trajectory $\vv{X}^{i}_{truth}(t)$ (green), and the up-scaled baseline trajectory $\vv{U}_4\vv{x}^{i}_{base}(t)$ (blue) at $t = 100\tau$  averaged over 20 different intial conditions $i$.}
\label{fig:spectrum}
\end{figure}

Figure~\ref{fig:spectrum} shows the power spectral density of the $\theta$ field after 100 time steps for the Ground Truth ($\vv{X}_{truth}(t)$), the ML-PDE trajectory ($\vv{X}_{ml}(t)$) and the up-scaled baseline ($\vv{U}_4\vv{x}_{base}(t)$).

\begin{figure*}
\centering
\includegraphics[width = 0.8\textwidth]{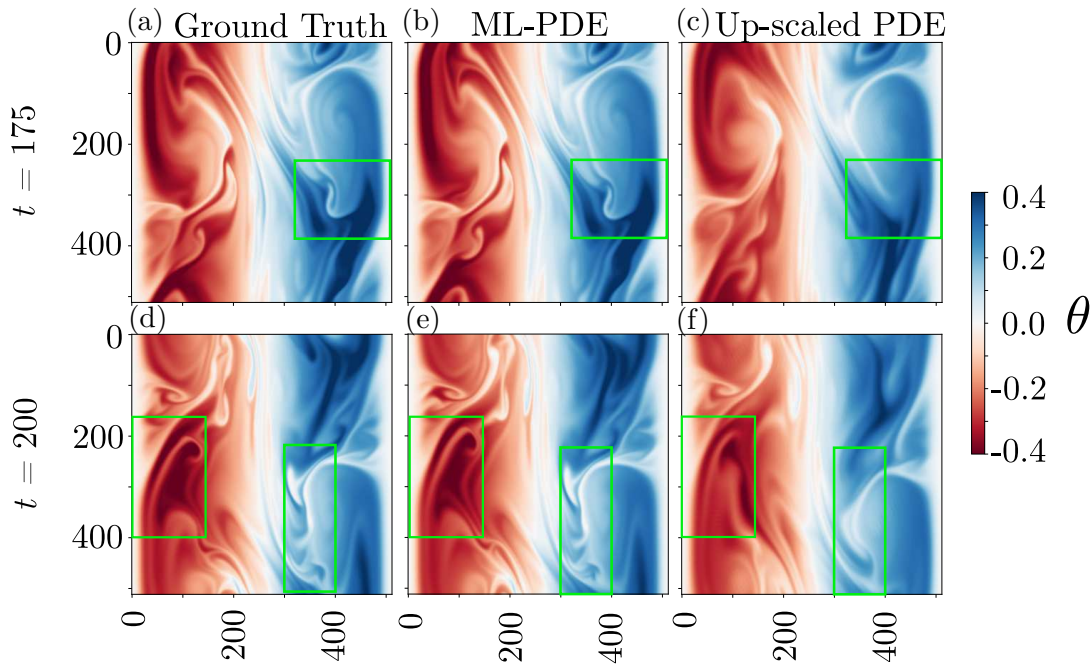}
\caption{(a), (d): Snapshot of the instantaneous temperature deviation ($\theta$) field from a ground truth simulation at time-steps $t = 175 \tau$ and $t = 200\tau$ respectively. (b), (e): The instantaneous $\theta$ field snapshot from the ML-PDE simulation. (c), (f): The instantaneous $\theta$ field snapshot from a baseline PDE simulation at $(128, 128)$ resolution up-scaled using the up-scaling operator $\vv{U}_4$. Interesting visually perceptible differences in the solution snapshots have been highlighted with a green box.}
\label{fig:snapshot}
\end{figure*}

In Figure~\ref{fig:snapshot}, we present a snapshot of the $\theta$ variable at two different instants of time, $t = 175\tau$ and $t = 200\tau$ as generated by the ML-PDE solver, the up-scaled baseline low-resolution PDE solver along with the ground truth for comparison. We observe significant, visually perceptible differences between the baseline solution and the ground truth whereas the ML-PDE solution has retains greater fidelity to the ground truth solution.

\section{\label{sec:conclusions} Conclusions}

We have introduced a novel ML-PDE hybrid architecture that can be used to effectively enhance the accuracy of a low-resolution solution of a complex application in incompressible fluid dynamics. The ML-PDE strategy features the use of a neural network to model the errors that accumulate over short time intervals between a coarse-grid evolution of the target PDE system and a solution of the system at much higher resolution. The resulting ML model is used to generate periodic corrections to a coarse-grid solution of the PDE system in a way that attempts to account for the missing fine scales.

We demonstrated our ML-PDE hybrid architecture using a canonical two-dimensional Rayleigh-B{\' e}nard Convection (RBC) system operating in a regime that exhibits a range of turbulent fluctuations with significant energy in high wave number modes. The ML-assisted coarse-grid evolution resulted in corrected solution trajectories that were consistent with the solutions computed at a much higher resolution in space and time.

While this demonstration example was carried out for a two-dimensional, quasi-steady flow, the approach can likely be generalized to more complex systems, and has the potential for pushing the envelope of high Reynolds number simulations across a variety of application areas. We acknowledge several important limitations of our initial implementation reported here.  First, three-dimensional turbulence problems of considerably more general interest will be dramatically more challenging to address, in terms of both complexity and required computational resources.  However, by the same arguments, we are optimistic that there is an opportunity for considerable benefit of our approach in that context.  We note that there is scope for further improving the architecture by introducing physics-based constraints on the output of the neural network, adversarial training of the neural network~\cite{goodfellow2014generative} and exploring different neural network architectures. For the two-dimensional example here, our approach is based on modeling fine scale features of the quasi-steady solution, and cannot in its present form capture the transient behaviour. Finally, we expect to explore the generalizeability of this modeling approach for the 2D RBC problem across varying Reynolds numbers, domain geometry and boundary conditions in future work. We hope that our work will motivate further research in all of these directions.

\section{Code and Data}
The code and data required to reproduce the results presented in the paper will be available at https://github.com/jdppthk/ML-PDE upon publication.

\section{Acknowledgements}
We would like to acknowledge the critical formative contributions to this work by Adrian Albert, who unfortunately passed before this manuscript was completed. We would also like to thank Prabhat, Peter Harrington and Wahid Bhimji for helpful comments and discussions throughout. This research used resources of the National Energy Research Scientific Computing Center (NERSC), a U.S. Department of Energy Office of Science User Facility operated under Contract No. DE-AC02-05CH11231.

\providecommand{\noopsort}[1]{}\providecommand{\singleletter}[1]{#1}%

\begin{appendices}
\section{Neural Network Architecture and Training\label{appendix:NN}}
The network follows the standard UNet design~\cite{ronneberger2015u} with a contracting path and an expanding path. The contracting path consists of $7$ convolution layers with kernel size$=4$ and stride$=2$, the input is padded in order to make the strided convolutions output exactly half of the input spatial size. The expanding path consists of $7$  transposed convolution layers with kernel size$=4$ and stride$=2$, again padding is chosen to make the output exactly double the input spatial size. To make this auto-encoder a UNet, the output of each convoultion layer from the contracting path is concatenated to the input of the corresponding layer in the expanding path, for example the output of the first layer is concatenated with the output of the penultimate layer to form the input of the very last layer of the network. Table~\ref{table:arch} provides model more details on the architecture.

The network was trained with {\tt apex.optimizers.FusedAdam}~\cite{apex}, a variant of Stochastic Gradient Descent optimizers, with batch-size$=30$ and learning-rate$=0.00039$ with a ReduceOnPlateau scheduler, the latter reduces the learning rate by a factor of 5x when the validation loss plateaus. We used L1 Norm as a loss between the predicted output and the ground truth, the target fields were multiplied by a factor of $10$ which we found to accelerate the training. The output fields are un-scaled before they are used in ML-PDE framework during inference.

\begin{table}[h]
\begin{tabular}{lccc}
\hline
 & \textbf{Output shape}  & \textbf{No. of params.} \\ \hline
Input variables     & 4 $\times$ 512 $\times$ 512   & - \\ \hline
ConvBlock &  128 $\times$ 256 $\times$ 256 & 9k \\
ConvBlock &  256 $\times$ 128 $\times$ 128 & 525k \\
ConvBlock &  512 $\times$ 64 $\times$ 64 & 2.1M \\
ConvBlock &  1024 $\times$ 32 $\times$ 32 & 8.4M \\
ConvBlock &  1024 $\times$ 16 $\times$ 16 & 16.8M \\
ConvBlock &  1024 $\times$ 8 $\times$ 8 & 16.8M \\
ConvBlock &  1024  $\times$ 4 $\times$ 4 & 16.8M \\ \hline
TConvBlock &  1024 $\times$ 8 $\times$ 8 & 16.8M \\
TConvBlock &  1024 $\times$ 16 $\times$ 16 & 33.5M \\
TConvBlock &  1024 $\times$ 32 $\times$ 32 & 33.5M \\
TConvBlock &  512 $\times$ 64 $\times$ 64 & 16.8M \\
TConvBlock &  256 $\times$ 128 $\times$ 128 & 4.2M \\
TConvBlock &  128 $\times$ 256 $\times$ 256 & 1M \\
TConv &  4 $\times$ 512 $\times$ 512 & 16k \\ \hline
\multicolumn{2}{l}{Total trainable parameters}      & \textbf{167M}    \\ \hline
\end{tabular}
\caption{Neural network architecture. Each ConvBlock consists of a Conv2D followed by a BatchNorm and LeakyReLU layers. Each TConvBlock consists of a TransponsedConv2D followed by a BatchNorm and ReLU layers. Last layer is a single TransposedConv2D layer followed by a $\tanh$ activation. Output shape is (channel $\times N_{x}$  $\times N_{y}$). Note that the output shape is illustrated for the $(512 \times 512)$ fields used during inference, the training was done on $(256 \times 256)$ as explained in section~\ref{sec:unet}.}
\label{table:arch}
\end{table}
\end{appendices}

\end{document}